\documentclass[a4paper]{jpconf}
\usepackage{graphicx}
\usepackage{amsmath}
\usepackage{slashed}
\usepackage{bbm}

\newcommand{\be}{\begin{equation}}
\newcommand{\ee}{\end{equation}}
\newcommand{\bea}{\begin{eqnarray}}
\newcommand{\eea}{\end{eqnarray}}
\newcommand{\nn}{\nonumber}
\DeclareMathOperator{\diag}{diag}

\newcommand{\gsim}{
\mathrel{\hbox{\rlap{\hbox{\lower4pt\hbox{$\sim$}}}\hbox{$>$}}}}

\begin{document}
\title{Non-minimally flavour violating dark matter}

\author{Monika Blanke}

\address{CERN Theory Division, CH-1211 Geneva 23, Switzerland\vspace{1mm}\\
Institut f\"ur Theoretische Teilchenphysik, Karlsruhe Institute of Technology,
Engesserstra{\ss}e 7,\\ D-76128 Karlsruhe, Germany\vspace{1mm}\\
Institut f\"ur Kernphysik, 
Karlsruhe Institute of Technology,
Hermann-von-Helmholtz-Platz 1,\\ D-76344 Eggenstein-Leopoldshafen, Germany}

\ead{monika.blanke@kit.edu}

\begin{abstract}
Flavour symmetries provide an appealing mechanism to stabilize the dark matter particle. I present a simple model of quark flavoured dark matter that goes beyond the framework of minimal flavour violation. I discuss the phenomenological implications for direct and indirect dark matter detection experiments, high energy collider searches as well as flavour violating precision data.
\end{abstract}

\section{Flavour symmetries in the dark sector}

While astrophysical and cosmological observations provide convincing evidence for the existence of dark matter (DM), little is known about its particle properties. Besides its gravitational interactions, appealing arguments exist that the DM particles should interact weakly with the Standard Model (SM). The reason for this belief is referred to as the ``WIMP miracle'': a DM particle with weak scale mass, $m_\text{DM}\sim\mathcal{O}(10^2\,\text{GeV})$ and weak coupling automatically generates the right DM relic abundance. The WIMP hypothesis receives further support from the electroweak (EW) naturalness problem, requiring an extension of the SM at the TeV scale in order to stabilize the (EW) symmetry breaking scale.

Adopting the WIMP hypothesis many properties of the DM particle are still unknown. It has to be neutral under strong and electromagnetic interactions, yet its transformation under EW symmetry can be non-trivial. Indeed the WIMP hypothesis seems to suggest a coupling to the weak gauge bosons. Nevertheless it is equally justified to introduce some new type of interactions through which an EW singlet DM particle couples to the SM. 

Non-trivial symmetry transformation properties of DM, both for global and local groups, are possible and worthwhile to be studied. In this article, summarizing the results of \cite{Agrawal:2014aoa}, we focus on a particular type of global symmetry, namely a flavour group.

The idea of flavoured DM is not new, see e.\,g.\ \cite{Kile:2013ola} for a review of early studies. However most studies so far assumed that the DM is charged under the SM flavour group and its interactions satisfy the Minimal Flavour Violation (MFV)  \cite{Buras:2000dm,D'Ambrosio:2002ex} hypothesis, so that the SM Yukawa couplings $Y_{u,d}$ remain the only source of flavour breaking. This approach has several benefits: apart from avoiding the stringent constraints from flavour changing neutral current (FCNC) processes, it can be shown \cite{Batell:2011tc} that the MFV principle implies the stability of DM, provided the latter transforms under certain representations of the flavour group. 

The drawback of the MFV assumption is that no interesting effects in flavour violating observables can be expected, thus closing a potential discovery channel of flavoured DM. Therefore in \cite{Agrawal:2014aoa} we studied the possibility of non-minimally flavour violating DM. The results of that paper have been presented at the DISCRETE 2014 conference and are summarized in the present article.

\section{Dark Minimal Flavour Violation and its implications}

\subsection{The Dark Minimal Flavour Violation hypothesis}
The Dark Minimal Flavour Violation (DMFV) hypothesis \cite{Agrawal:2014aoa} is a conceptually straightforward extension of the well-known MFV principle to the DM sector. In complete analogy to the SM matter content we assume that the DM $\chi$ transforms under the fundamental representation of a new global flavour symmetry $U(3)_\chi$. The full flavour symmetry group\footnote{in the absence of right-handed neutrinos} then reads
\begin{equation}
G_F = U(3)_q \times U(3)_u \times U(3)_d \times U(3)_\ell \times U(3)_e \times U(3)_\chi\,.
\end{equation}
The crucial ingredient for DMFV is then that this global flavour symmetry is broken \emph{only} by the SM Yukawa couplings $Y_u$, $Y_d$, $Y_e$, and the DM coupling $\lambda$ to the SM fermions. Depending on which type of SM fermion the DM field $\chi$ couples to, and depending on the Dirac structure of the coupling, different classes of DMFV models can be obtained. 

We are thus left with a huge variety of models with interesting and quite distinct phenomenological implications. It is important to note that while DMFV is conceptually close to the MFV ansatz, the phenomenology of DMFV models differs vastly from the MFV ones. This is due to the new coupling matrix $\lambda$ whose structure is unrelated to the SM Yukawa couplings, and upon which no hierarchical structure is imposed. 

\subsection{A minimal model}

For concreteness, following \cite{Agrawal:2014aoa}, let us focus on a concrete class of DMFV models in what follows. We introduce DM as a Dirac fermion $\chi$ that transforms as a triplet under the global flavour symmetry $U(3)_\chi$, and that couples to right-handed down-type quarks via a scalar mediator $\phi$.  The field $\phi$ is singlet under the flavour group $G_F$, however it transforms as 
\begin{equation}
\phi \sim (\mathbf{3}, \mathbf{1})_{-1/3}
\end{equation}
under the SM gauge group $SU(3)_C\times SU(2)_L\times U(1)_Y$. The DM field $\chi$ is a gauge singlet. In order to render the coupling
\begin{equation}
\lambda \bar d_R \chi \phi
\end{equation}
formally invariant under the flavour group $G_F$, $\lambda$ is promoted to a spurion field which transforms as triplet under $U(3)_d$ and as anti-triplet under $U(3)_\chi$. Table \ref{tab:representations} summarizes the transformation properties of all matter and spurion fields present in this DMFV framework.

\begin{table}[!h]
  \begin{center}
    \begin{tabular}{|c||ccc|cccc|}
      \hline
      & $SU(3)_c$ & $SU(2)_L$ & $U(1)_Y$ & $U(3)_q$ & $U(3)_u$ & $U(3)_d$ & $U(3)_\chi$ \\ \hline\hline
      $q_L$ &  {\bf 3} & {\bf 2} & 1/6 & {\bf 3} & {\bf 1} & {\bf 1} & {\bf 1} \\
      $u_R$ & {\bf 3} & {\bf 1} & 2/3 & {\bf 1} & {\bf 3} & {\bf 1} & {\bf 1} \\ 
      $d_R$ & {\bf 3} & {\bf 1} & -1/3 & {\bf 1} & {\bf 1} & {\bf 3} & {\bf 1} \\ \hline
      $\ell_L$ &  {\bf 1} & {\bf 2} & -1/2 & {\bf 1} & {\bf 1} & {\bf 1} & {\bf 1} \\
      $e_R$ &  {\bf 1} & {\bf 1} & -1 & {\bf 1} & {\bf 1} & {\bf 1} & {\bf 1} \\ \hline
      $H$ &  {\bf 1} & {\bf 2} & 1/2 & {\bf 1} & {\bf 1} & {\bf 1} & {\bf 1} \\ \hline
      $\phi$ &  {\bf 3} & {\bf 1} & -1/3 & {\bf 1} & {\bf 1} & {\bf 1} & {\bf 1} \\ 
      $\chi_L$ &  {\bf 1} & {\bf 1} & 0 & {\bf 1} & {\bf 1} & {\bf 1} & {\bf 3} \\ 
      $\chi_R$ &  {\bf 1} & {\bf 1} & 0 & {\bf 1} & {\bf 1} & {\bf 1} & {\bf 3} \\ \hline\hline
      $Y_u$ &  {\bf 1} & {\bf 1} & 0 & {\bf 3} & \boldmath{${\bar 3}$} & {\bf 1} & {\bf 1} \\ 
      $Y_d$ &  {\bf 1} & {\bf 1} & 0 & {\bf 3} & {\bf 1} & \boldmath{${\bar 3}$}  & {\bf 1} \\ 
      $\lambda$  &  {\bf 1} & {\bf 1} & 0 & {\bf 1}  & {\bf 1} & {\bf 3} & \boldmath{${\bar 3}$}  \\ \hline
    \end{tabular}
  \end{center}
  \caption{Symmetry transformation properties of the minimal DMFV matter content and the Yukawa spurions.\label{tab:representations}}
\end{table}

The minimal DMFV (mDMFV) model coupling DM to down-type quarks and containing only the matter content summarized in table \ref{tab:representations} is fully described by the renormalisable Lagrangian
\begin{eqnarray}
  {\cal L}&=& \mathcal{L}_\text{SM}+
  i \bar \chi \slashed{\partial} \chi
  - m_{\chi}  \bar \chi \chi  - (\lambda_{ij} \bar {d_{R}}_i \chi_j \phi + {\rm h.c.}) \nn \\
  && \qquad\,
  +
  (D_{\mu} \phi)^{\dagger} (D^{\mu} \phi) - m_{\phi}^2 \phi^{\dagger} \phi
  +\lambda_{H \phi}\, \phi^{\dagger} \phi\, H^{\dagger} H
  +\lambda_{\phi\phi}\, \phi^{\dagger} \phi\, \phi^{\dagger} \phi \,.\label{eq:Lagrangian}
\end{eqnarray}
Here $i,j=1,2,3$ are flavour indices. We stress that the $3\times 3$ coupling matrix $\lambda$ constitutes a new source of flavour violation with respect to MFV models. Before we turn to summarizing the phenomenological implications of this model, we recapitulate a number of direct implications of the DMFV ansatz. We stress that these implications follow from symmetry principles alone and therefore hold beyond the minimal model.

\subsection{Implications of DMFV}

The attentive reader might have already noticed that the DM mass term $m_\chi$ in \eqref{eq:Lagrangian} does not carry flavour indices. Indeed the DMFV principle ensures the universality of masses for the three DM flavours. Yet this statement strictly speaking is true only at the leading order in the DMFV spurion expansion. Including the leading non-trivial contribution the mass matrix becomes
\begin{align}\label{eq:DMFVmass}
  m_{ij} =   m_{\chi} (\mathbbm{1}
  +\eta\, \lambda^\dagger\lambda+ \cdots
  )_{ij}\,.
\end{align}
The flavour violating spurion $\lambda$ thus introduces a mass splitting between the various DM flavours, and the size of this splitting is intimately tied to the size of deviations from a universal DM coupling to quarks. Furthermore DMFV ensures that the DM mass matrix is aligned with the coupling $\lambda$ in the $U(3)_\chi$ flavour space. 

The coefficient $\eta$ in \eqref{eq:DMFVmass} is a free parameter of the low energy theory, and can be predicted only in a UV-complete flavour model. If the correction to $m_\chi$ is generated at the tree level, it is expected to be of $\mathcal{O}(1)$. Tree level contributions might be absent, depending on the details of the UV completion. A mass splitting of the form $\lambda^\dagger\lambda$ is however unavoidably generated at the one loop level, leaving us with the constraint $|\eta|\gsim\mathcal{O}(10^{-2})$. In our phenomenological analysis we will be agnostic about the size of $\eta$. In order to ensure perturbativity of the DMFV expansion we will however impose that the mass splittings generated by $\eta\, \lambda^\dagger\lambda$ are at most 30\%.

Another welcome implication of the DMFV assumption is the reduction of free parameters in the flavour sector. The $U(3)_\chi$ flavour symmetry can be used to remove a number of non-physical parameters from the coupling matrix $\lambda$. The latter can then be parametrised in terms of six real parameters and three CP violating phases as follows. We split $\lambda$ into a unitary matrix $U_\lambda$ and a positive diagonal matrix $D_\lambda$,
\be
\lambda = U_{\lambda} D_{\lambda} \,,
\ee
and then parametrise $U_\lambda$ in terms of three mixing angles $s^\lambda_{12,13,23}$ and three phases $\delta^\lambda_{12,13,23}$ as in \cite{Blanke:2006xr}. For $D_\lambda$ we use two parametrisations that can easily translated into each other
\be\label{eq:Dlambda}
D_\lambda \equiv \diag(D_{\lambda,11},D_{\lambda,22},D_{\lambda,33})=
\lambda_0\cdot \mathbbm{1}+\diag(\lambda_1,\lambda_2,-(\lambda_1+\lambda_2))\,.
\ee

The most interesting consequence of the DMFV ansatz, as far as the DM phenomenology is concerned, is the implied stability of the lightest state in the DM sector. An explicit analysis (see appendix B of \cite{Agrawal:2014aoa}) shows that the flavour symmetry $G_F$ broken only by the SM Yukawa couplings and the new coupling $\lambda$, together with the unbroken $SU(3)_C$ gauge symmetry, contain an unbroken $\mathbbm{Z}_3$ subgroup under which only $\chi$ and $\phi$ transform non-trivially.
Consequently neither $\phi$ nor any of the flavours of $\chi$ can decay into SM particles only, and the lightest $\chi$ flavour is predicted to be stable. At the same time, none of these states can be singly produced at the LHC, and their decay will give rise to missing energy signatures, similar to the ones of supersymmetric squarks.

Last but not least the DMFV hypothesis also ensures that the results of our flavour analysis, while obtained within the mDMFV model, hold beyond the minimal particle content. We will get back to this point in the following section.

\section{Phenomenology of the minimal DMFV model}

\subsection{Flavour constraints}\label{sec:flavour}

\begin{figure}[h!]
\centering
\includegraphics[width=0.25\textwidth]{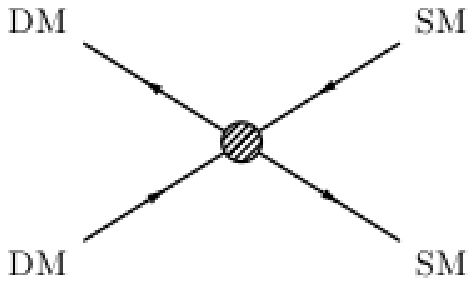}\qquad\qquad
\includegraphics[width=0.25\textwidth]{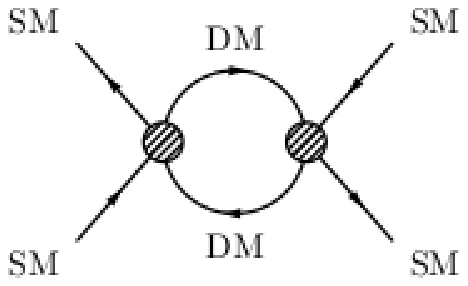}
\caption{Schematic diagrams contributing to experimental constraints on flavoured DM \cite{Agrawal:2014aoa}.}
\label{fig:DMloop}
\end{figure}

The coupling $\lambda$ of DM to SM quarks is a new source of flavour violation and therefore generates new contributions to FCNC observables. This gives rise to an exciting new discovery opportunity, in addition to the usual channels in direct and indirect detection experiments and collider searches. See figure \ref{fig:DMloop} for schematic diagrams. At the same time however we also have to deal with the stringent constraints from well measured flavour violating decays.

\begin{figure}[h!]
\centering
\includegraphics[width=0.33\textwidth]{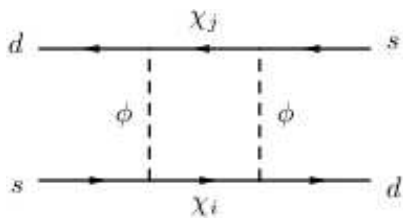}
\caption{New contribution to $K^0-\bar K^0$ mixing in the mDMFV model \cite{Agrawal:2014aoa}.}
\label{fig:kkbar}
\end{figure}

Like in many extensions of the SM, the most restrictive constraints on the mDMFV parameter space arise from meson anti-meson mixing observables. New contributions are generated by the exchange of the new particles $\phi$ and $\chi$ at the loop level, see figure \ref{fig:kkbar}. The full expression for the mixing amplitude can be found in  \cite{Agrawal:2014aoa}. While the exact expression for the loop function has to obtained from an explicit calculation, the flavour violating structure can be obtained from the DMFV principle by taking the leading term in the spurion expansion. We find
\be\label{eq:M12}
M_{12}^{K,\text{new}} \sim
  \left((\lambda\lambda^\dagger)_{sd}\right)^2 { F(x) }\,,\quad 
M_{12}^{d,\text{new}} \sim
  \left((\lambda\lambda^\dagger)_{bd}\right)^2 { F(x) }\,,\quad 
M_{12}^{s,\text{new}} \sim
  \left((\lambda\lambda^\dagger)_{bs}\right)^2 { F(x) }\,,
\ee
where $F(x)$ is the flavour universal loop function.

\begin{figure}[htp]
\centering{
\includegraphics[width=.325\textwidth]{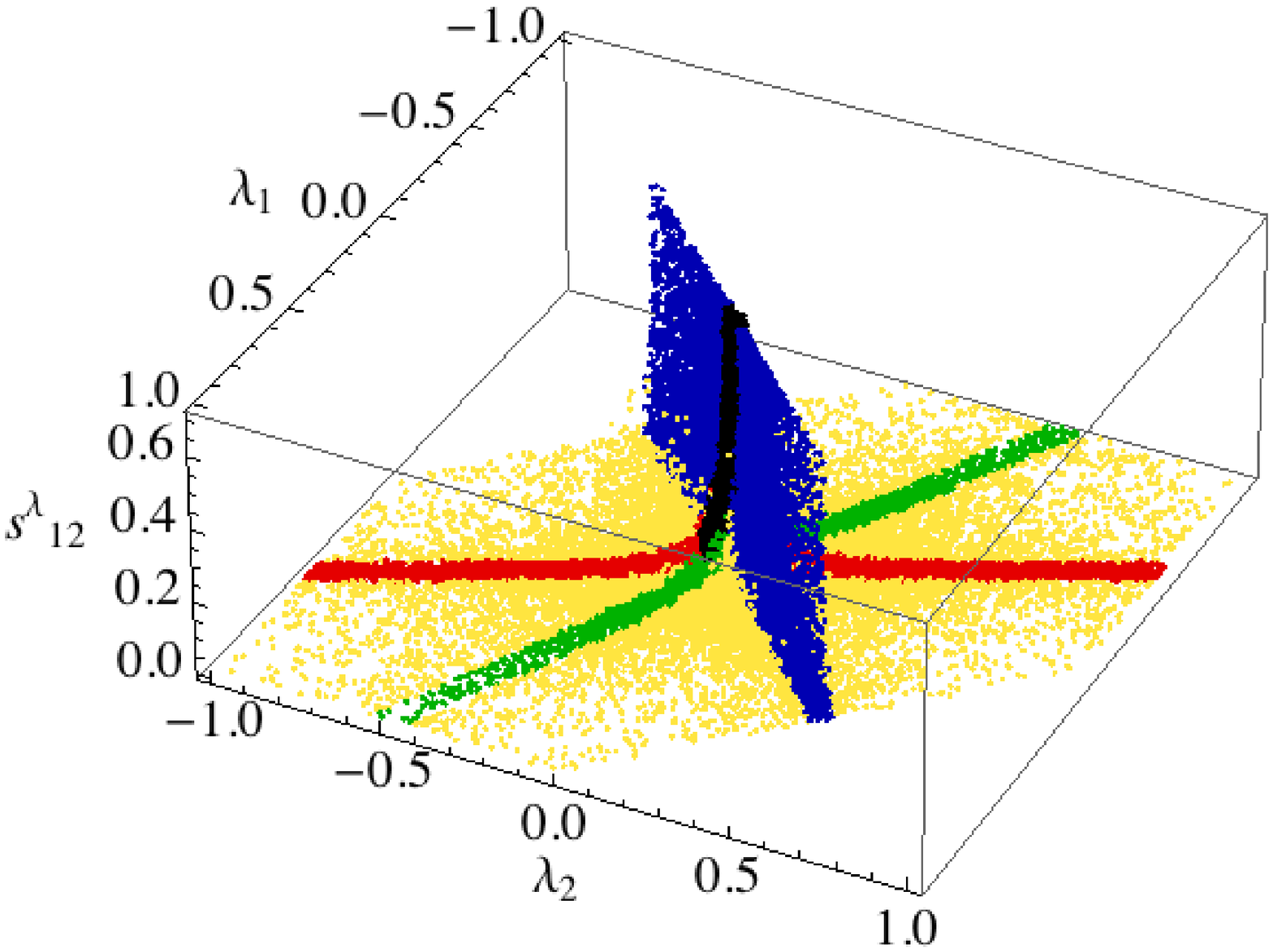}
\hfill
\includegraphics[width=.325\textwidth]{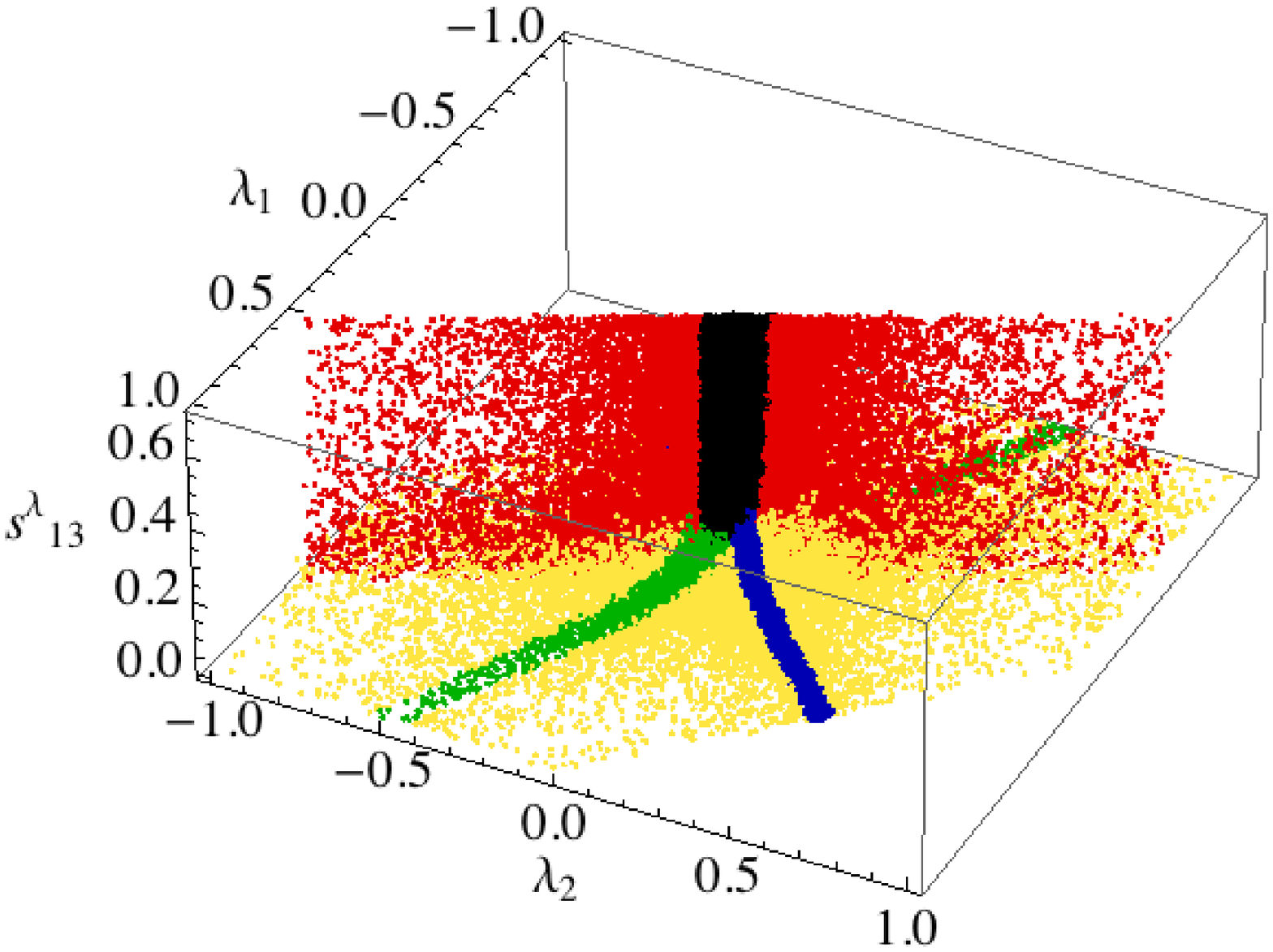}
\hfill
\includegraphics[width=.325\textwidth]{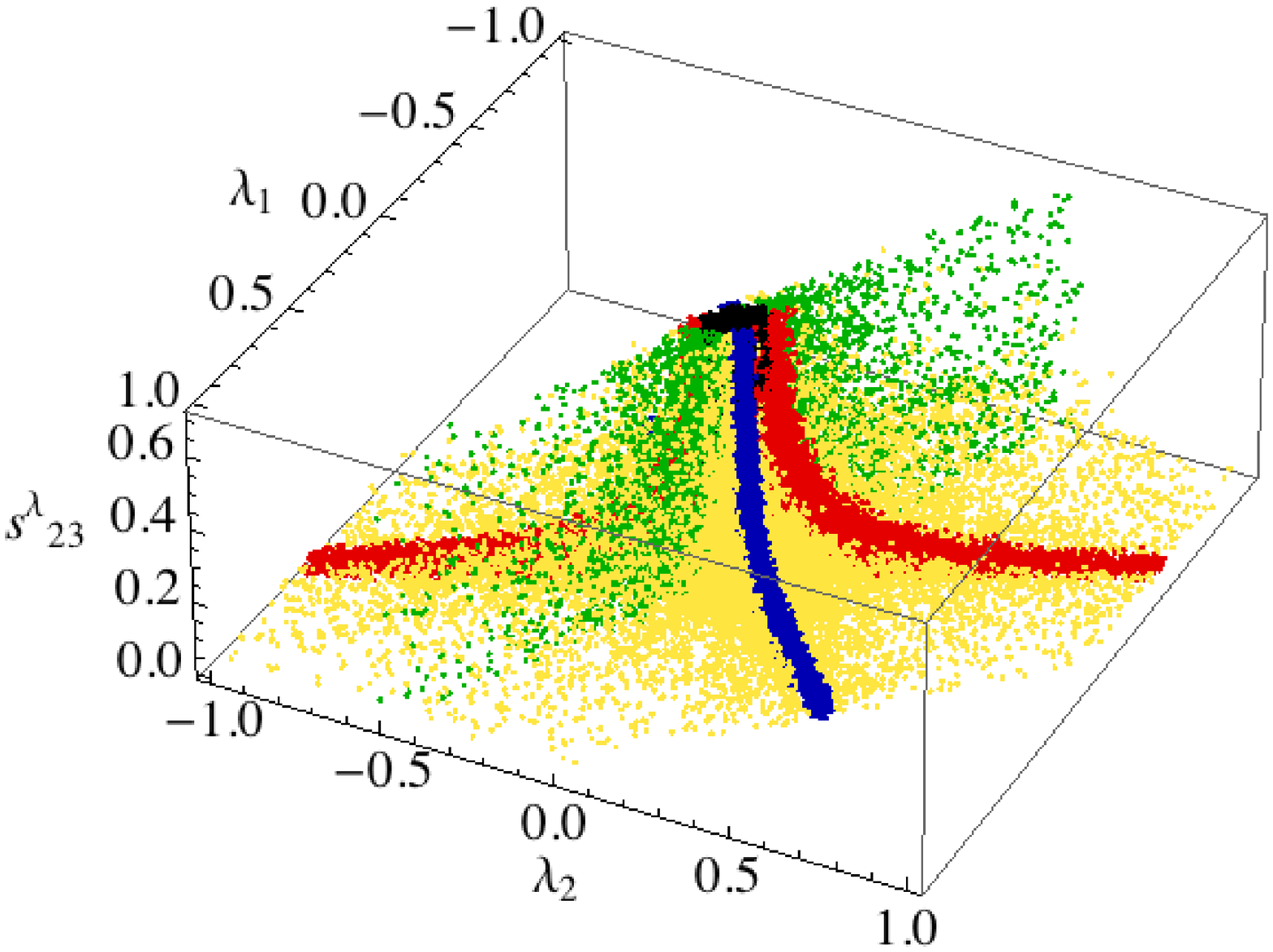}}
\caption{\label{fig:scenarios}Scenarios for the structure of $\lambda$: Universality (black), 12-degeneracy (blue), 13-degeneracy (red), 23-degeneracy (green), 
small mixing (yellow) \cite{Agrawal:2014aoa}.}
\end{figure}

Equation \eqref{eq:M12} shows that the mDMFV contributions to the $\Delta F =2 $ mixing amplitudes -- and in fact to all FCNC transitions -- are driven by the off-diagonal elements of $ \lambda\lambda^\dagger$. A number of scenarios for the structure of $\lambda$ can thus be derived, and confirmed in a numerical parameter scan (c.\,f.\ figure \ref{fig:scenarios}).

\begin{enumerate}
\item {\bf universality scenario} (black): $\lambda_1 \simeq \lambda_2  \simeq 0 $

In this case $\lambda \simeq U_\lambda \cdot \lambda_0$ so that $\lambda \lambda^\dagger \simeq \lambda_0^2\cdot\mathbbm{1}$. 

\item {\bf 12-degeneracy} (blue): $\lambda_1 \simeq \lambda_2$

If the first two generations of DM fermions are quasi-degenerate, then  the mixing angle $s_{12}^\lambda$ can be generic while $s^\lambda_{13,23}$ have to be small. This can be understood by taking the limit $\lambda_1=\lambda_2$, in which case the mixing angle $s_{12}^\lambda$ becomes non-physical.

\item {\bf 13-degeneracy} (red): $\lambda_2 \simeq -2\lambda_1$

In the case $\lambda_2 \simeq -2\lambda_1$ the first and third DM flavour are quasi-degenerate, and consequently $s_{13}^\lambda$ is unconstrained. In order to suppress the remaining flavour violating effects both $s^\lambda_{12}$ and $s^\lambda_{23}$ have to be small.

\item {\bf 23-degeneracy} (green): $\lambda_2 \simeq -1/2\lambda_1$

Finally if $\lambda_2 \simeq -1/2\lambda_1$, the second and third DM flavour are quasi degenerate. Consequently the mixing angle $s^\lambda_{23}$ is arbitrary, while $s^\lambda_{12}$ and $s^{\lambda}_{13}$ have to be small.

\item
{\bf small mixing scenario} (yellow): arbitrary $D_\lambda$

Last but not least if $D_\lambda$ does not exhibit any degeneracies, then FCNC effects have to be suppressed by the smallness of all three mixing angles $s_{12}^\lambda \simeq s_{13}^\lambda \simeq s_{23}^\lambda \simeq 0$. This scenario, shown by the yellow points in figure \ref{fig:scenarios}, corresponds to a diagonal but non-degenerate coupling matrix $\lambda$. 
\end{enumerate}

It is important to note that equation \eqref{eq:M12} is a direct consequence of the DMFV ansatz, without further assumption on the particle content of the model. The above scenarios will therefore also hold in non-minimal models with additional particles, as long as the coupling $\lambda$ remains the only new source of flavour violation.

While meson anti-meson mixing observables place severe bounds on the mDMFV parameter space, the new contributions to rare $K$ and $B$ decays turn out to be small. One the one hand we do not need to worry about the constraints from well-measured decays like $B\to X_s\gamma$, yet on the other hand at least within the minimal model we can neither explain the anomaly in the $B\to K^+\mu^+\mu^-$ data nor can we hope for a spectacular new physics signature in rare decays that have not yet been measured, like $K\to \pi\nu\bar\nu$.

\subsection{Dark matter phenomenology}

In order to proceed with the DM phenomenology of the mDMFV model we have to make a choice for the mass spectrum in the DM sector, i.\,e.\ which of the $\chi$ flavours is the lightest. We identify the various flavours by which quark flavour $d$, $s$ or $b$ they couple to dominantly. From a theoretical point of view, all three choices for the lightest DM flavour are equally well justified. The case of $d$-flavoured DM is however disfavoured by the lack of signal in the direct detection experiments, at least if we assume DM to be a thermal relic. $b$- and $s$-flavoured DM are similar to each other, as far as direct detection experiments are concerned. However $b$-flavoured DM can explain the recently observed $\gamma$ ray excess from the galactic centre \cite{Agrawal:2014una}. It also gives rise to interesting $b$-jet signatures at the LHC. We therefore restrict our attention to this case, as also done in \cite{Agrawal:2014aoa}. 

While it would be straightforward to circumvent this constraint from a model building point of view, we assume for simplicity that the DM relic abundance is generated by thermal freeze-out. Depending on the mass splitting in the DM sector different scenarios have to be considered. When their masses
are nearly degenerate, then all DM flavours are present during
freeze-out and can be treated together. Otherwise only the lightest
flavour of DM remains in the thermal bath. The relic abundance condition then constrains the couplings of all DM flavours present in the thermal bath, as a function of the DM mass and the mediator mass.

\begin{figure}[h!]
\centering
  \includegraphics[width=0.25\textwidth]{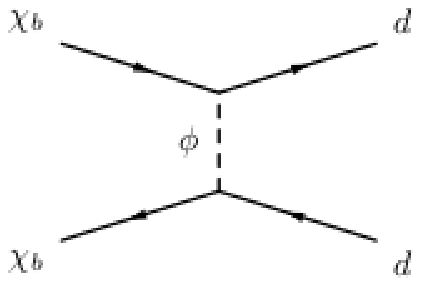}\qquad
  \includegraphics[width=0.25\textwidth]{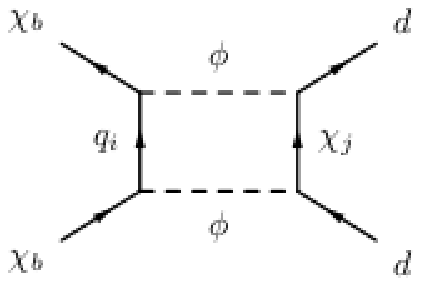}\qquad
  \includegraphics[width=0.25\textwidth]{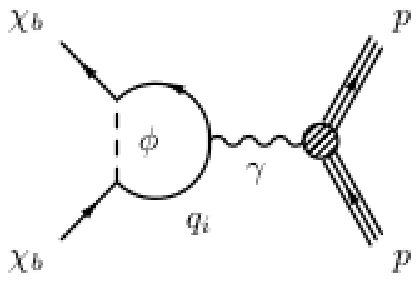}
  \caption{Diagrams contributing to WIMP-nucleon scattering in the mDMFV model \cite{Agrawal:2014aoa}.}
  \label{fig:dd}
\end{figure}

Significant constraints on the parameter space arise from the LUX experiment, searching for DM scattering off nuclei. In the mDMFV model with $b$-flavoured DM the relevant diagrams contributing to WIMP-nucleon scattering are shown in figure \ref{fig:dd}.

In order to efficiently suppress the tree level contribution, the mixing between the first and third DM generation, $s^\lambda_{13}$, has to be small. Non-negligible effects arise however from the one-loop box and penguin diagrams. These can not be suppressed by choosing small flavour mixing angles, as they are present even in the flavour conserving case. However they interfere destructively with each other thanks to an overall minus sign of the leading logarithmic term in the penguin diagram. This cancellation becomes effective if $D_{\lambda,11}$, the first generation DM coupling to quarks, lies in a certain range.

\begin{figure}[h!]
\centering
\includegraphics[width=.4\textwidth]{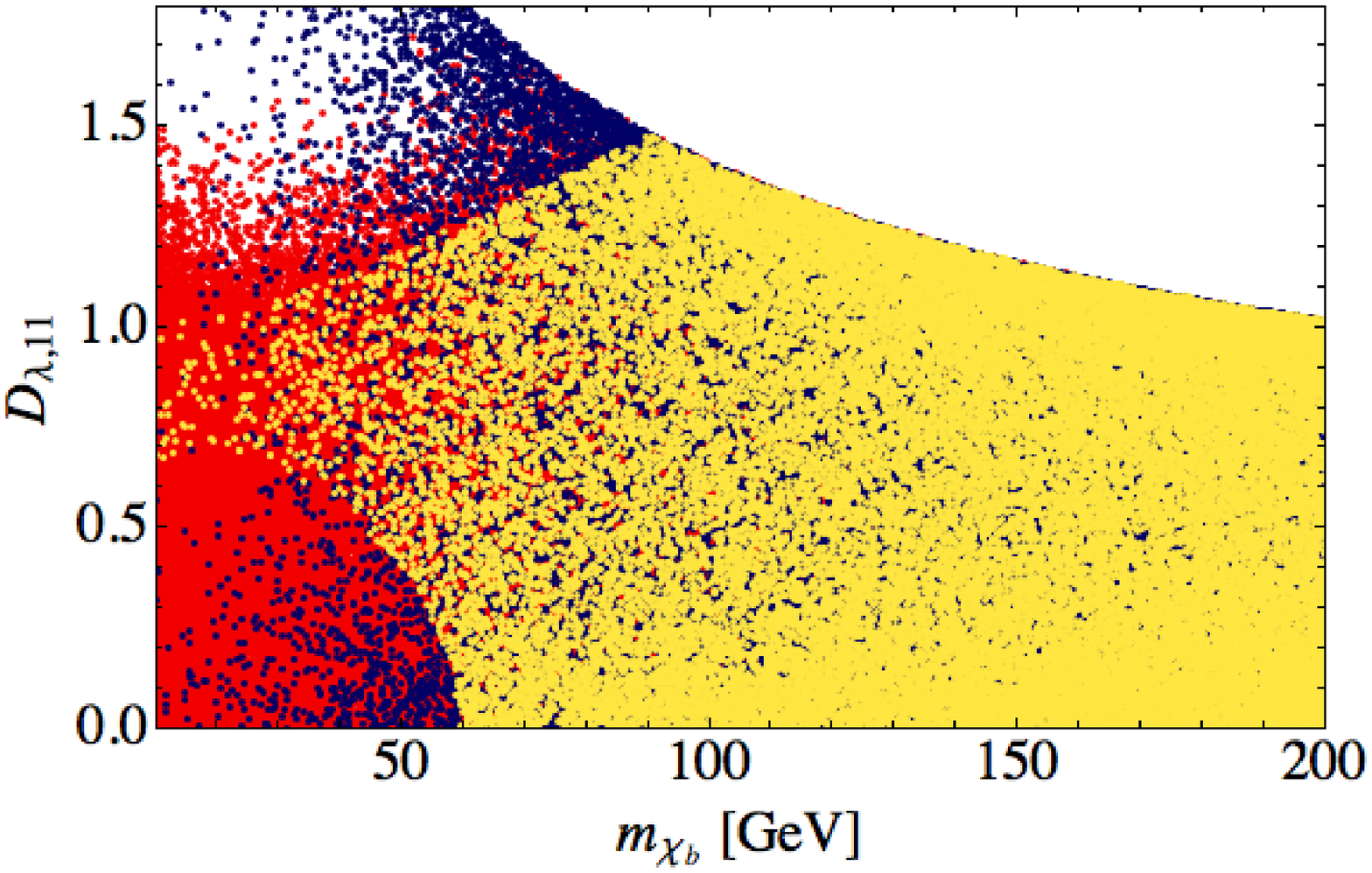}\qquad
\includegraphics[width=.4\textwidth]{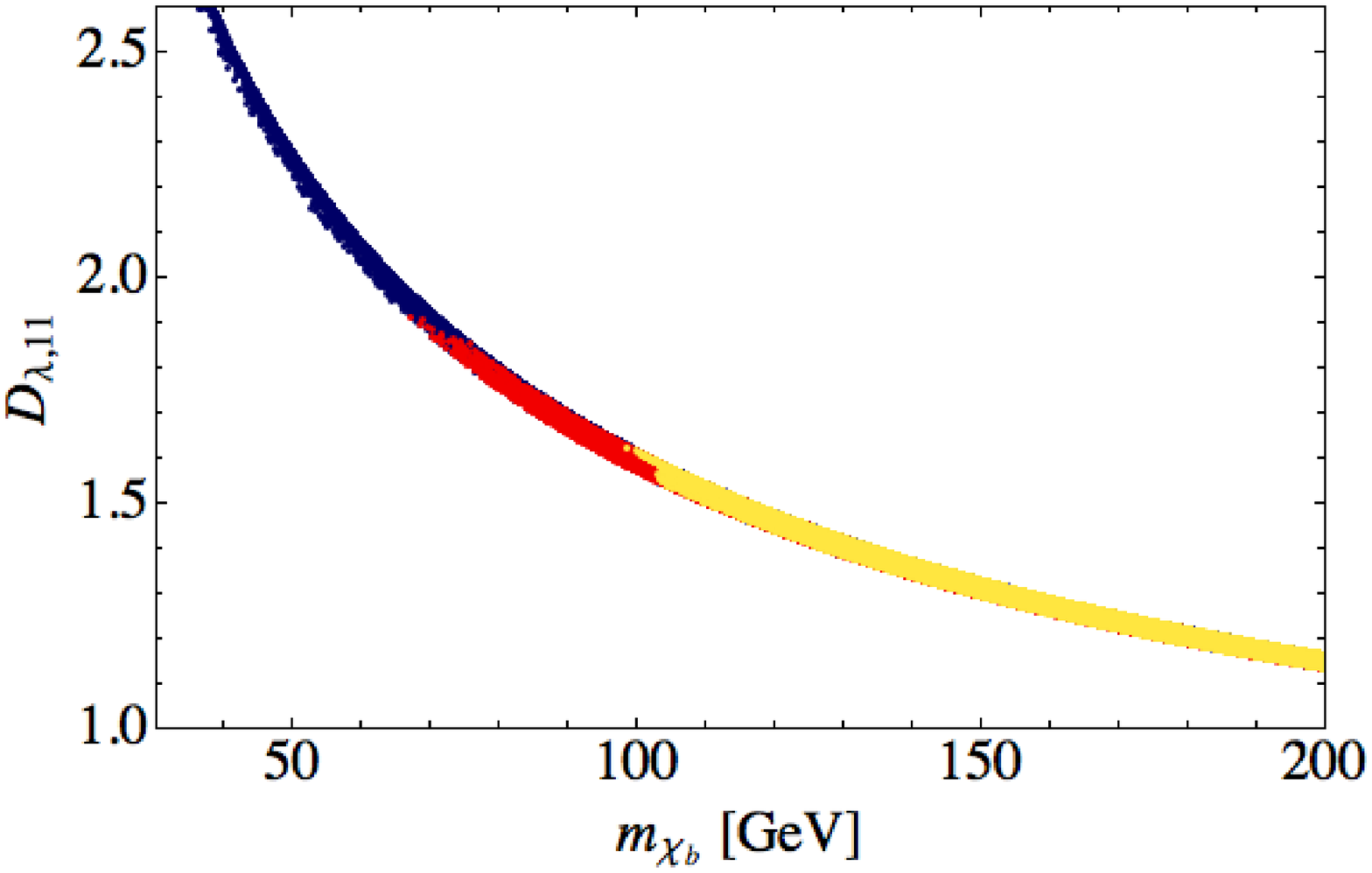}
\caption{\label{fig:mchi-D11}First generation coupling $D_{\lambda,11}$ as functions of the DM mass $m_{\chi_b}$ for significant mass splitting $m_{\chi_{d,s}}>1.1m_{\chi_b}$ scenario (left), and in the $13$-degeneracy scenario $m_{\chi_{d}} \simeq m_{\chi_b}$ (right). The red points satisfy the bound from LUX, while the blue points satisfy the $\Delta F=2$ constraints. For the yellow points both LUX and $\Delta F =2$ constraints are imposed \cite{Agrawal:2014aoa}.}
\end{figure}

This effect can be observed in figure \ref{fig:mchi-D11}. The region covered by yellow points is the one that remains valid after LUX and flavour constraints have been taken into account. The left panel shows the allowed range for the coupling $D_{\lambda,11}$ as a function of the DM mass $m_{\chi_b}$ in the case of a single DM flavour freeze-out. For small DM mass where the LUX bound is strongest, we can clearly see that only a small range for the coupling $D_{\lambda,11}\sim 1$ is allowed. Interestingly this constraint is not visible from the red points, where only the LUX bound but not the flavour constraints has been imposed.

The restriction to a certain range of $D_{\lambda,11}$ has a striking consequence in the case where the first and third DM generations are quasi-degenerate. In this case the relic abundance constraint and the LUX bound become incompatible below DM masses below $m_{\chi_b}\sim 100\,\text{GeV}$. Lower DM masses are therefore excluded in the $13$-degeneracy scenario, as can be seen from the right panel of figure \ref{fig:mchi-D11}.

\subsection{Bounds from LHC searches}

While the new particles of the DMFV framework have not yet been the subject of experimental analyses, interesting and complementary constraints on the mDMFV parameter space can be obtained from existing searches for new particles at the LHC.  Similarly to supersymmetric models with conserved R-parity, also the mDMFV states have to be pair-produced and lead to missing energy signatures.

\begin{figure}[!h]
  \begin{center}
    \includegraphics[width=0.45\textwidth]{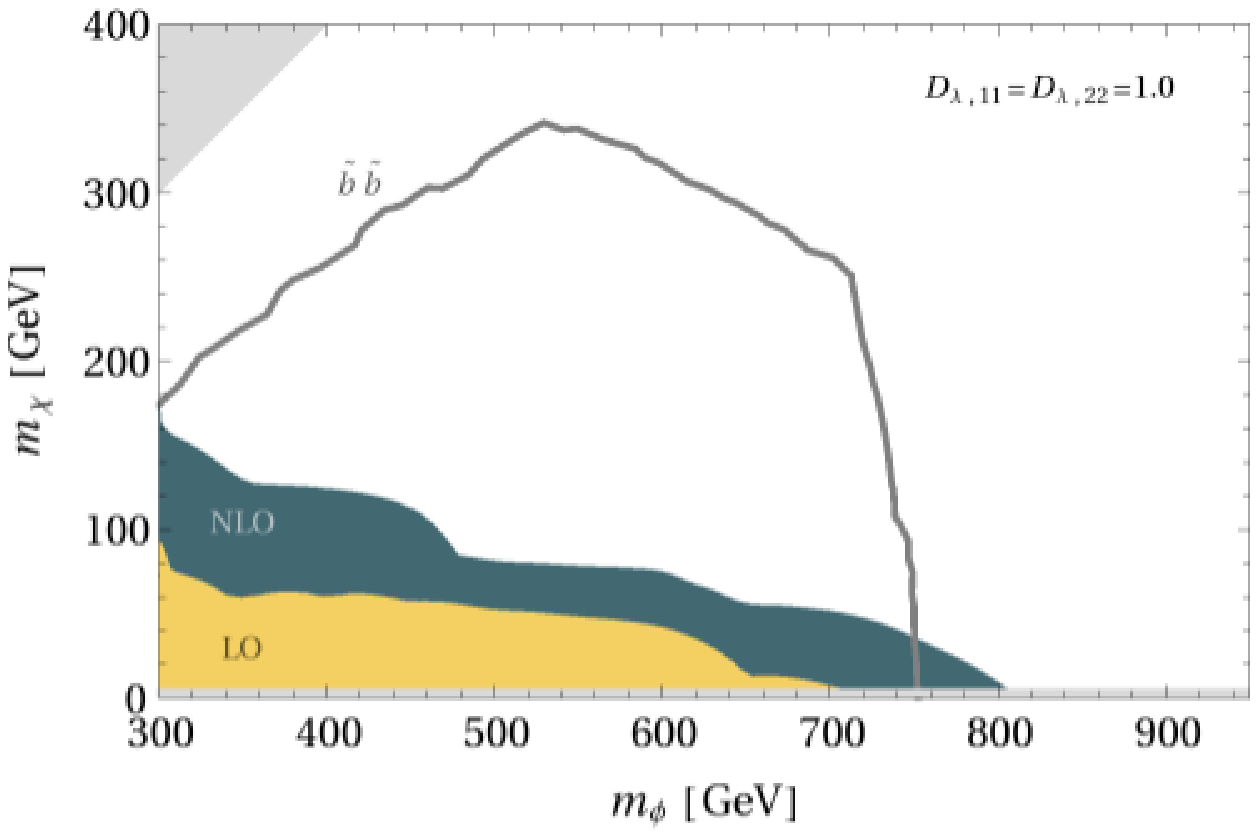}
    \quad
    \includegraphics[width=0.45\textwidth]{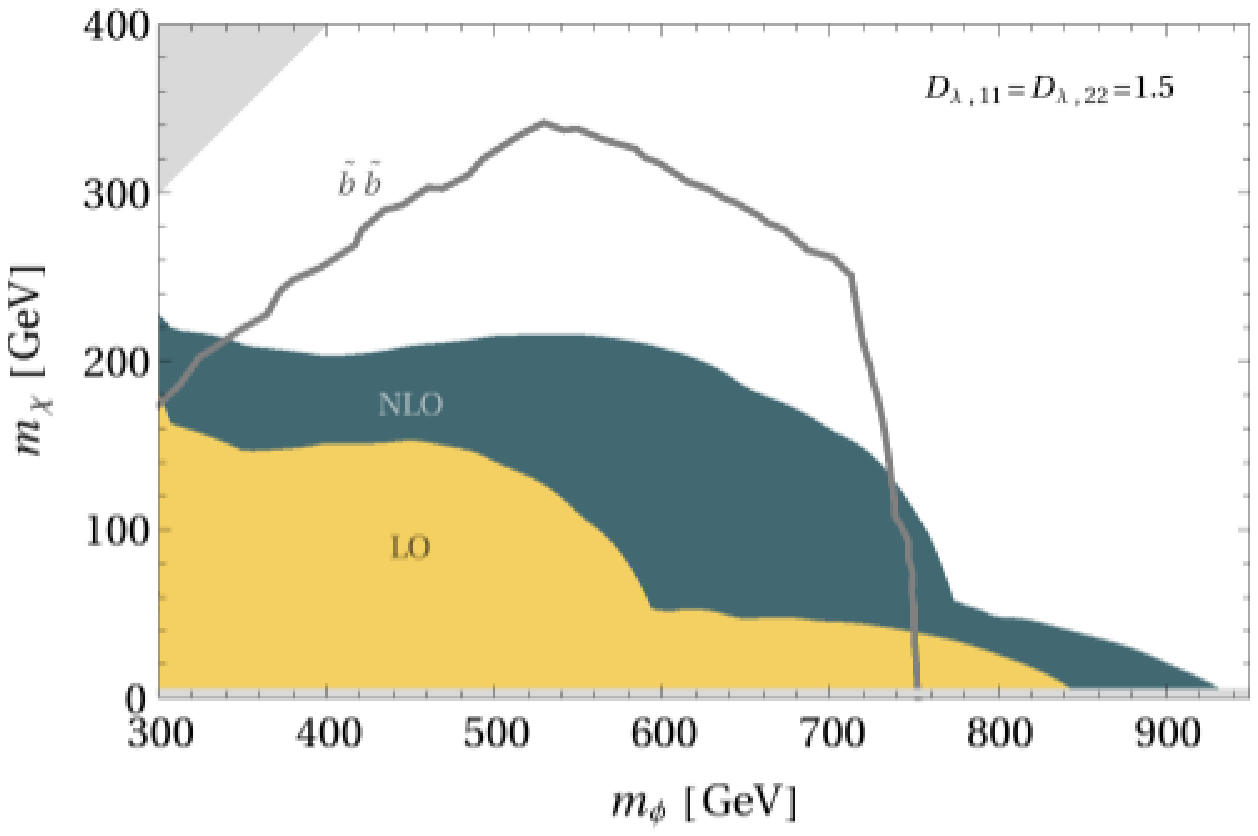}
  \end{center}
  \caption{Limits on the mDMFV model from CMS squark searches \cite{Agrawal:2014aoa}.}
  \label{fig:sbottoms1}
\end{figure}

We can therefore obtain a bound on the masses $m_\phi$ and $m_\chi$ from the experimental searches for squark pair production. Figure \ref{fig:sbottoms1} shows the limits on the mDMFV model obtained from the CMS 19.5   fb$^{-1}$ sbottom~\cite{CMS:2014nia} and
  squark~\cite{Chatrchyan:2014lfa} searches. To this end both the modified production cross section and the branching ratios to the various DM flavours have been taken into account. Since the heavier DM flavours' decay leads to additional soft particles in the final state, which are not considered in the experimental analysis, all three flavours have been treated as missing energy.  The DM coupling to the $b$ quark, $D_{\lambda,33}$, has been fixed everywhere by the corresponding relic abundance constraint, and all mixing angles are set to zero for simplicity. We can see that mediator masses up to $m_\phi\sim 850\,\text{GeV}$ are excluded for light DM masses $m_\chi$, however the bounds become weaker for heavier $\chi$.

\begin{figure}[!h]
  \begin{center}

    \includegraphics[width=0.45\textwidth]{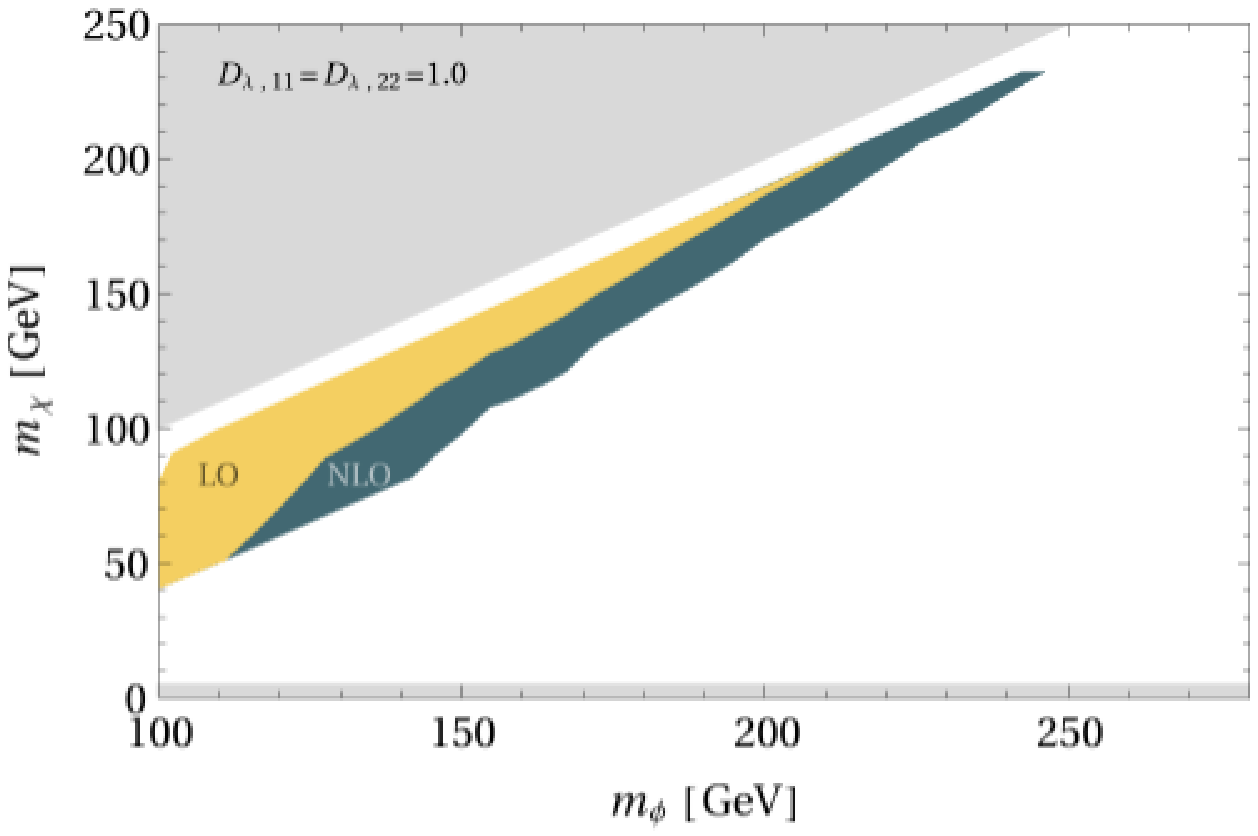}
    \includegraphics[width=0.45\textwidth]{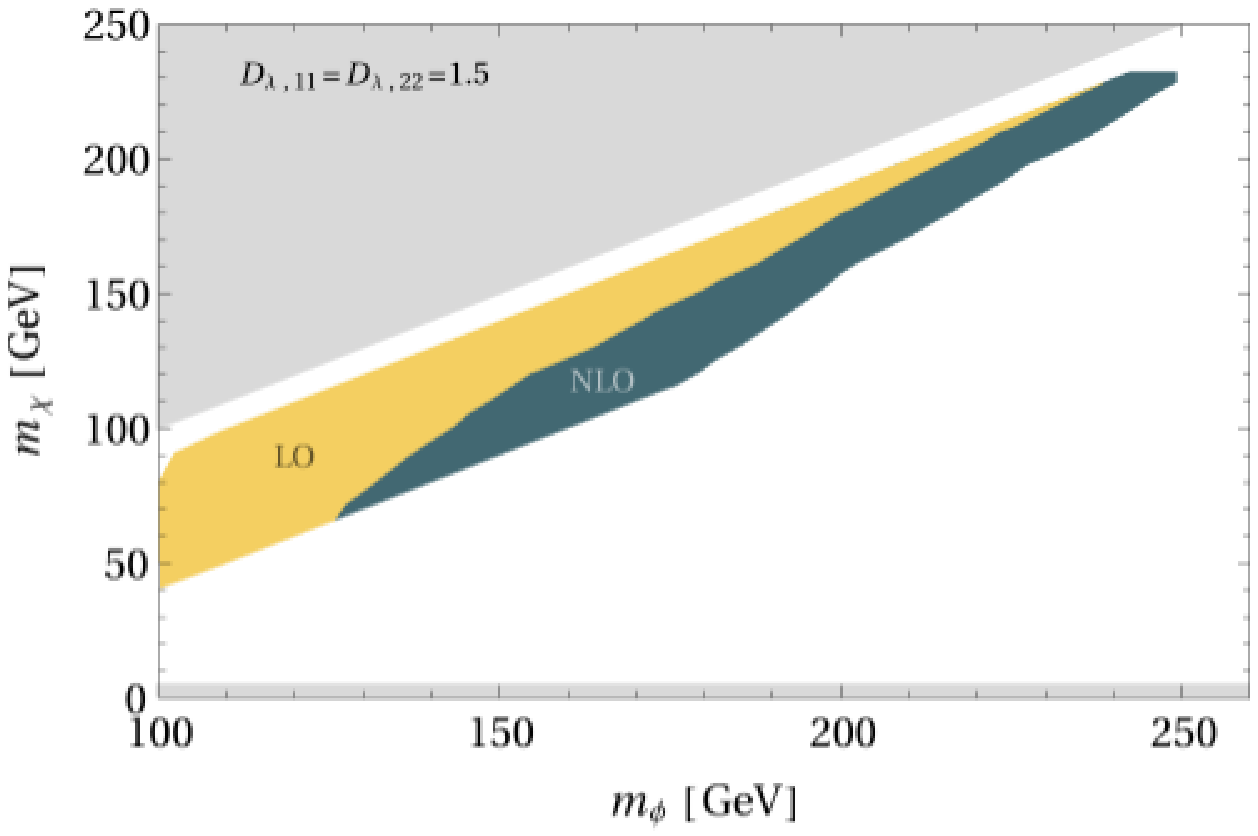}
  \end{center}
  \caption{Limits on the mDMFV model from the CMS 19.7
  fb$^{-1}$ search~\cite{CMS:2014yma} for stops decaying to a
  charm and a neutralino,
  using the monojet + MET final state \cite{Agrawal:2014aoa}. 
  }
  \label{fig:monojets}
\end{figure}

The CMS search for light stops decaying into a charm and a neutralino \cite{CMS:2014yma} places a constraint on a different region of mDMFV parameter space, see figure \ref{fig:monojets}. An excluded region appears when the mediator $\phi$ and the DM $m_\chi$ are close to degenerate, so that the decay products of $\phi$ are soft. The constraint reaches up to $m_\phi \sim 250\,\text{GeV}$.

\begin{figure}[!h]
  \begin{center}
    \includegraphics[width=0.45\textwidth]{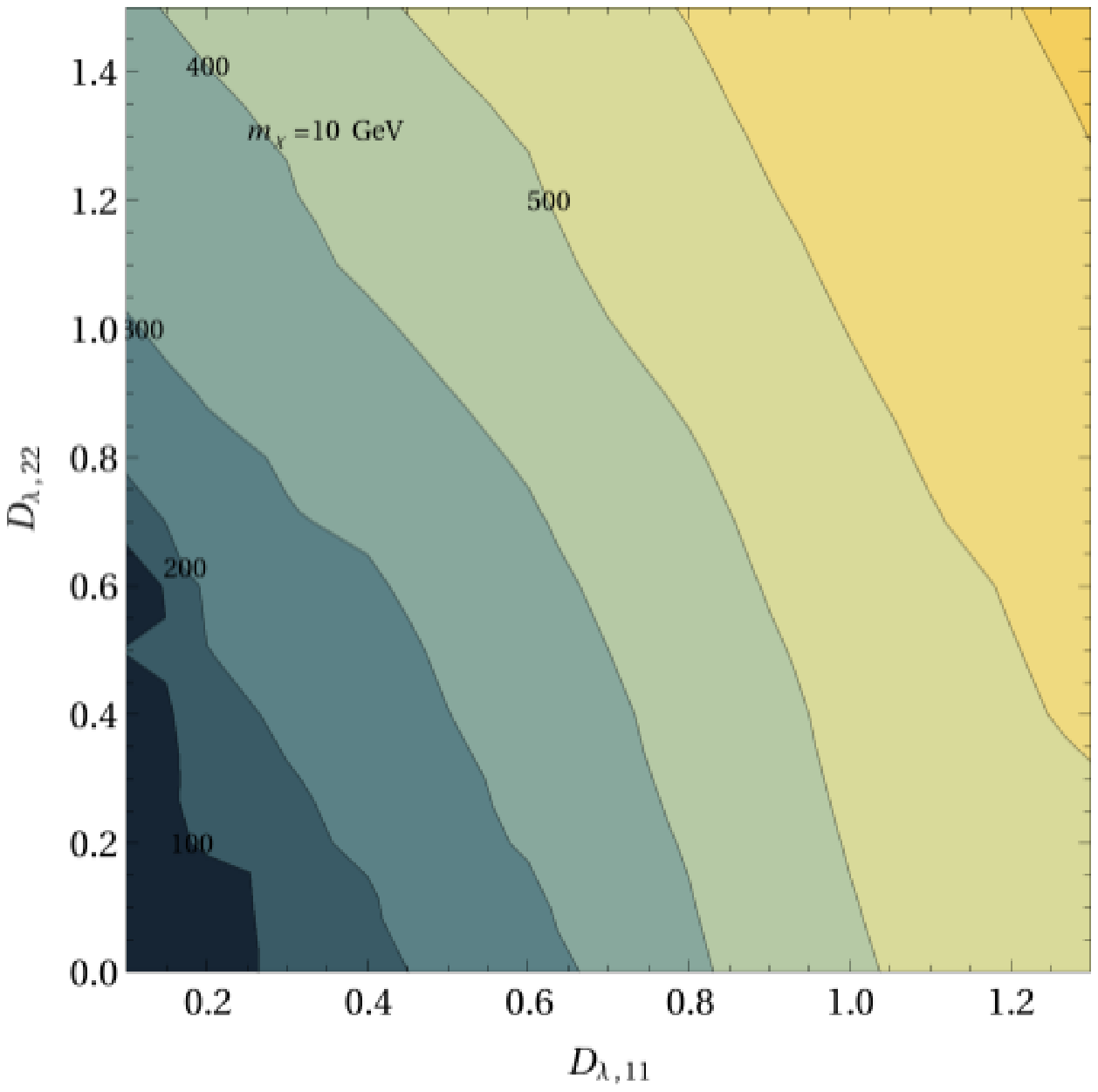}
    \qquad
    \includegraphics[width=0.45\textwidth]{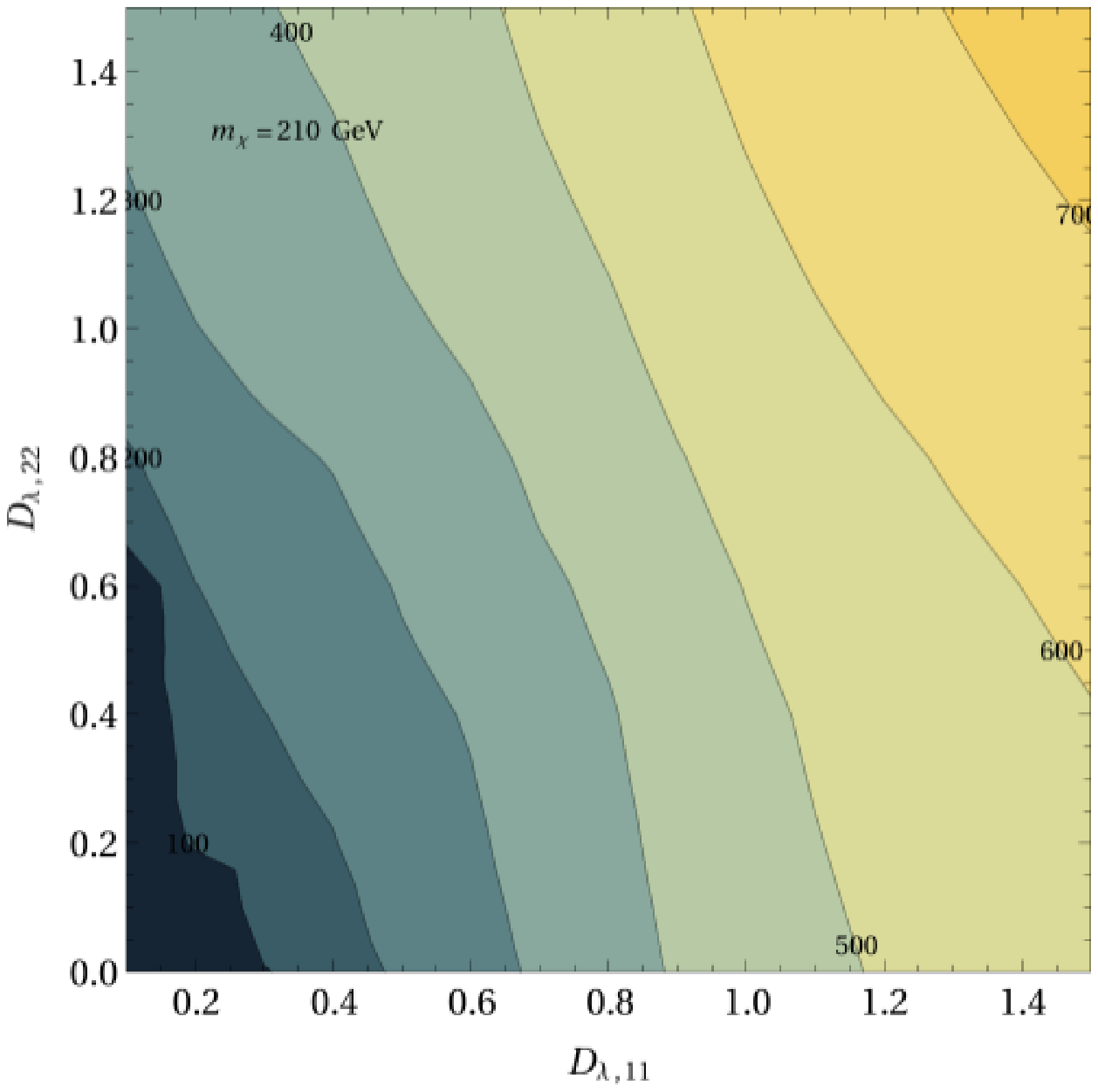}
  \end{center}
  \caption{Limits on the mDMFV model in terms of mass of $\phi$ (in
  GeV) from the CMS monojet
  search~\cite{CMS:rwa} \cite{Agrawal:2014aoa}. }
  \label{fig:monojetscoup}
\end{figure}

Last but not least monojet searches that place constraints on effective DM interactions are also sensitive to flavoured DM. In figure \ref{fig:monojetscoup} we show the result of reinterpreting the CMS monojet
  search~\cite{CMS:rwa} in terms of the mDMFV model. Note that in this case the obtained constraints on the mediator mass $m_\phi$ are rather insensitive to the DM mass $m_\chi$.

\section{Summary}

Flavoured DM is an exciting scenario on the astroparticle physics playground, that has interesting implications for flavour physics, collider physics, and direct DM searches, with a non-trivial interplay between the various areas. In this article we have reviewed a simplified model of quark flavoured DM beyond the MFV hypothesis, in which the flavour violating DM coupling to quarks is the only new source of flavour violation. The mDMFV model realises only one out of many possibilities to realize the flavoured DM paradigm in a simplified model scenario. Distinct yet equally interesting implications can be expected in other implementations that deserve further study.

\end{document}